\documentclass[aps,twocolumn,amsmath,amssymb,floatfix]{revtex4-1}
\usepackage{graphicx}
\usepackage{dcolumn}
\usepackage{bm}
\usepackage{amsfonts}
\usepackage{graphicx}
\usepackage{bm}
\usepackage{amsmath}
\usepackage{amssymb}

\begin{document}
\title{Stacking dependence of carrier-interactions in multilayer graphene systems}
\author{Yunsu Jang$^{1}$}
\author{E. H. Hwang$^{2}$}
\author{A. H. MacDonald$^{3}$}
\author{Hongki Min$^{1}$}
\email{hmin@snu.ac.kr}
\affiliation{$^1$ Department of Physics and Astronomy, Seoul National University, Seoul 151-747, Korea}
\affiliation{$^2$ SKKU Advanced Institute of Nanotechnology and Department of Physics, Sungkyunkwan University, Suwon 440-746, Korea}
\affiliation{$^{3}$Department of Physics, University of Texas at Austin, Austin TX 78712}
\date{\today}

\begin{abstract}
We identify qualitative trends in the stacking sequence dependence of carrier-carrier interaction
phenomena in multilayer graphene.  Our theory is based on a new approach which explicitly 
exhibits the important role in interaction phenomena of the momentum-direction dependent intersite phases
determined by the stacking sequence.  Using this method, we calculate and compare the self-energies, density--density response functions, collective modes, 
and ground-state energies of several different few layer graphene systems.  The influence of 
electron--electron interactions on important electronic properties  
can be understood in terms of competition between intraband exchange, interband exchange and correlation contributions
that vary systematically with stacking arrangement.
\end{abstract}

\maketitle

{\em Introduction} --- 
Multilayer graphene has attracted considerable attention recently because of exotic chiral features in its electronic structure and the possibility of future electronic device applications\cite{CastroNeto2009,DasSarma2011,Basov2014,HassanRaza2012}.  
The band structure of a multilayer system is qualitatively 
dependent on its stacking sequence, opening up the possibility of engineering electronic properties by selecting a desired 
arrangement.  In this paper we use an approach in which momentum-direction dependent intersite phases
determined by the stacking sequence are explicitly exhibited to show that this qualitative dependence is inherited by carrier-carrier interaction phenomena.  

Because the number of $\pi$-bands in a multilayer graphene system is proportional to the number of layers, and because 
$\pi$-band wavefunctions are not isotropic in momentum space, accurate evaluation of physical quantities which 
require integrations over momentum space, for example quasiparticle energy spectra and density--density correlation functions, 
rapidly becomes more difficult as layer number increases.  
To mitigate this problem and to make the relationship between stacking arrangement and interactions more 
transparent, we introduce a momentum-direction dependent unitary transformation which 
makes the single-particle Hamiltonian isotropic.  
In addition to making accurate many-electron perturbation theory calculations practical for multilayer stacks,
this approach facilitates understanding of some qualitative trends in 
the stacking arrangement dependence of quasiparticle energy spectra, plasmon dispersion and damping, and carrier thermodynamic properties. 


{\em Rotational transformation of multilayer Dirac Hamiltonian} --- 
Our calculation is based on the minimal continuum model for multilayer graphene which retains only a 
Dirac model for hopping within each layer and only nearest-neighbor interlayer hopping.
Different stacking sequences are specified by different interlayer near-neighbor arrangements.
The Hamiltonians for these minimal models can be made isotropic by multiplying wavefunction components by stacking and 
momentum-direction dependent phase factors. To illustrate how this transformation works, 
we consider first the example of Bernal stacked bilayer graphene, in which one sublattice in the first layer (say 1B) is a 
near neighbor of the opposite sublattice in the second layer (say 2A). 
The Hamiltonian at finite wavevector ${\bm k}$ is then expressed in the (1A, 1B, 2A, 2B) basis as
\begin{equation}
\label{eq:hamiltonian_unbiased}
{\cal H}(\phi_{\bm{k}})=\left(
\begin{array}{cccc}
0 &\hbar v k e^{-i\phi_{\bm{k}}} &0 &0 \\
\hbar v k e^{i\phi_{\bm{k}}} &0 &t_{\perp} &0 \\
0 &t_{\perp} &0 &\hbar v k e^{-i\phi_{\bm{k}}}\\
0 &0 &\hbar v k e^{i\phi_{\bm{k}}} &0\\
\end{array}
\right),
\end{equation}
where $k=\sqrt{k_x^2+k_y^2}$, $\phi_{\bm{k}}=\arctan(k_y/k_x)$, $v$ is the bare Dirac velocity, which is related to the nearest-neighbor intralayer hopping
amplitude by $t= 2 \hbar v /\sqrt{3}a \sim 3$ eV ($a=0.246$ nm is the lattice constant), and $t_{\perp}\sim 0.1t$ is the nearest-neighbor interlayer hopping parameter. 
It is easy to see that the eigenvalues of this Hamiltonian are independent of 
$\phi_{\bm{k}}$ and that all eigenvalues 
satisfy 
$\Psi(\phi_{\bm{k}})=(c_{1\rm A},c_{1\rm B}e^{i\phi_{\bm{k}}},c_{2\rm A}e^{i\phi_{\bm{k}}},c_{2\rm B}e^{2i\phi_{\bm{k}}})^{\rm t}\equiv U(\phi_{\bm{k}}) \Psi(0)$, where 
the $\{c_i\}$ depend on $k$ only and can be obtained by diagonalizing $\cal{H}$ at $\phi_{\bm{k}}=0$.  
The locking between intersite phases and momentum direction in these spinors 
is reminiscent of the properties of spinors in chiral systems and will be referred to below as 
sublattice pseudospin chirality.  
The unitary operator $U(\phi)$ is a diagonal matrix
whose diagonal components $(1,e^{i\phi},e^{i\phi},e^{2i\phi})$ are determined by the bilayer stacking. 
The phase difference $e^{i\phi}$ between the ${1\rm A}$ and ${1\rm B}$ components of the wavefunction, and between 
the ${2\rm A}$ and ${2\rm B}$ components, comes from the monolayer-like intralayer coupling, whereas the zero phase difference 
between the ${1\rm B}$ and ${2\rm A}$ components comes from the momentum-independent interlayer coupling.
If we know the wavefunction at a specific angle, we can easily obtain the wavefunction at an arbitrary angle by attaching 
site-dependent phase factors determined 
by the stacking sequence.

\begin{figure}[htb]
\includegraphics[width=1\linewidth]{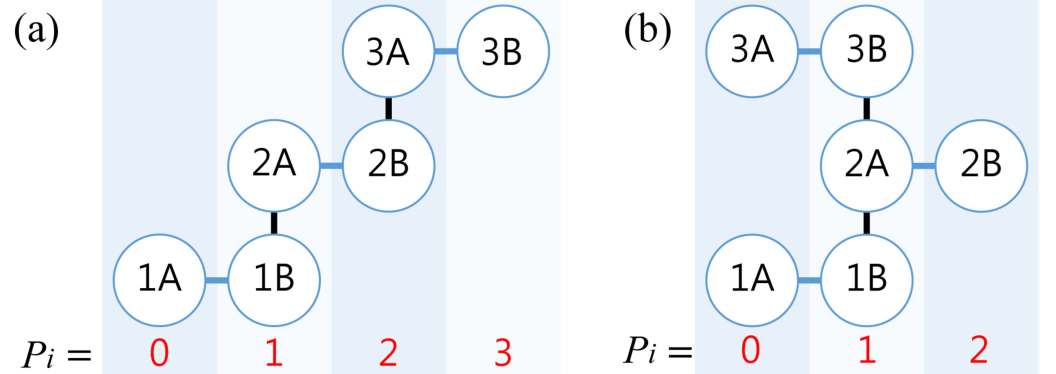}
\begin{ruledtabular}
\begin{tabular}{ccccccccc}
\,\,(c)\phantom{aaaa} & $P_{1A}$ & $P_{1B}$ & $P_{2A}$ & $P_{2B}$ & $P_{3A}$ & $P_{3B}$ & $P_{4A}$ & $P_{4B}$  \\
\hline
A&0&1&&&&&&\\
AB&0&1&1&2&&&&\\
ABC&0&1&1&2&2&3&&\\
ABA&0&1&1&2&0&1&&\\
ABCA&0&1&1&2&2&3&3&4\\
ABCB&0&1&1&2&2&3&1&2 \\
ABAB&0&1&1&2&0&1&1&2\\
ABAC&0&1&1&2&0&1&-1&0 \\
\end{tabular}
\end{ruledtabular}
\caption{
Stacking diagrams and phase factor chirality parameters $\{P_i\}$ for (a) ABC and (b) ABA graphene. (c) Phase factors for all stacking arrangements from
monolayers to tetralayers. We have chosen to set the phase factor of the sublattice 1A to zero. These results are for valley $K$.
For valley $K'$ the chirality parameters change sign.  
} 
\label{fig:phase_factor}
\end{figure}

We can easily generalize from the bilayer case to multilayer graphene with an arbitrary stacking order. 
Eigenstates at momentum orientation $\phi_{\bm{k}}$ satisfy 
\begin{equation}
\Psi(\phi_{\bm{k}})=(c_{1\rm A}e^{iP_{1\rm A} \phi_{\bm{k}}},c_{1\rm B} e^{iP_{1\rm B}\phi_{\bm{k}}},\cdots)^{\rm t}=U(\phi_{\bm{k}})\Psi(0),
\label{eq:wavefunction_rotated}
\end{equation}
where ${\cal H}(0)\Psi(0)=\varepsilon\Psi(0)$.  ${\cal H}(\phi_{\bm{k}})=U(\phi_{\bm{k}}){\cal H}(0)U^{-1}(\phi_{\bm{k}})$ has matrix 
elements
\begin{equation}
{\cal H}_{ij}(\phi_{\bm{k}})={\cal H}_{ij}(0) e^{i(P_i-P_j)\phi_{\bm{k}}},
\label{eq:matrix_element}
\end{equation}
and eigenvalues $\varepsilon$ that are independent of $\phi_{\bm{k}}$.  
By comparing the matrix elements in Eq.~(\ref{eq:matrix_element}) with those in the 
original Hamiltonian, we can determine the phase factor 
chirality parameters $\{P_i\}$. 
In general two sites connected by nearest-neighbor interlayer hopping have the same phase and within a layer 
$P_{\rm B} = P_{\rm A}+1$.
Using these two rules, $\{P_i\}$ is completely determined by the stacking sequence. 
Figure \ref{fig:phase_factor} illustrates their application in multilayer structures with up to four layers. 
We explain below how these band structure properties influence electron-electron interaction physics in multilayer graphene systems.

{\em Exchange self-energy} --- Our goal in this paper is to address interaction effects in moderate carrier density multilayer graphene systems, which are are weakly correlated two-dimensional 
Fermi liquids in which electron-electron interaction effects can be reliably addressed using perturbation theory.  
At leading order the electron self-energy is given by the unscreened exchange contribution:
\begin{equation}
\Sigma_{\rm ex}({\bm k},s)=-\sum_{s'}\int {d^2k' \over (2\pi)^2} V_{{\bm k}-{\bm k}'} f_{{s',\bm k'}} F_{\bm{k},\bm{k}'}^{s,s'} ,
\label{eq:exchange self-energy}
\end{equation}
where $f_{s,{\bm k}}$ is the Fermi function for the band $s$ and wavevector ${\bm k}$, 
$F_{\bm{k},\bm{k}'}^{s,s'}=\left|\left<s,{\bm k}|s',{\bm k'}\right>\right|^2$ is a wavefunction overlap factor, 
and $V_{q}=2\pi e^2 / \epsilon_{0} q$ is the two-dimensional Coulomb interaction.
(We note that the Coulomb interaction between layers with the layer separation $d$ is given by $V_{q}(d)=V_q e^{-qd}$. Due to the small layer separation we approximate $V_q(d) \approx V_q$ for the analytic calculations. The full numerical calculations using $V_q(d)$ do not change the results qualitatively.)
It is conventional to absorb the self-energy at the Dirac point ($\bm{k}=0$) in the absence of carriers into the 
zero of energy.  

To understand the consequences for interaction physics of multilayer wavefunction chiral properties, 
it is instructive to first consider the chiral two-dimensional electron system (C2DES) Hamiltonians\cite{min2008a} that provide a low-energy effective theory of multilayer graphene.
The Hamiltonian of a C2DES with the chirality index $J$ is  
\begin{equation}
{\cal H}_J(\bm{k})=t_{\perp}\left(
\begin{array}{cc}
0 & \left({\hbar v k e^{-i\phi_{\bm{k}}} \over t_{\perp}} \right)^J \\
\left({\hbar v k e^{i\phi_{\bm{k}}} \over t_{\perp}}\right)^J & 0 \\
\end{array}
\right),
\end{equation}
and yields eigenenergies $\varepsilon_{s,\bm{k}}=s t_{\perp} \left(\hbar v |\bm{k}| / t_{\perp}\right)^J$, and 
eigenspinors  $\left|s,\bm{k}\right> = \left(s,e^{i J\phi_{\bm{k}}}\right)^{\rm t}/\sqrt{2}$, 
where $s=\pm 1$ for positive and negative energy states respectively. 
For a C2DES with the chirality $J$, 
$F_{\bm{k},\bm{k}'}^{s,s'}={1\over 2}\left[1+ss'\cos J(\phi_{\bm{k}}-\phi_{\bm{k}'})\right]={1\over 2}\left(1+ss'{\bm n}_{\bm{k}}\cdot{\bm n}_{\bm{k}'}\right)$, 
where ${\bm n}_{\bm{k}}=(\cos J\phi_{\bm{k}},\sin J\phi_{\bm{k}})$ is the pseudospin direction at ${\bm k}$
characterized by the chirality index $J$. 
Note that the overlap factor $F_{\bm{k},\bm{k}'}^{s,s'}$ for a C2DES has the form of Heisenberg interactions between pseudospins with orientation $J\phi_{\bm{k}}$ and $J\phi_{\bm{k}'}$\cite{min2008b}.

\begin{figure}[htb]
\includegraphics[width=1\linewidth]{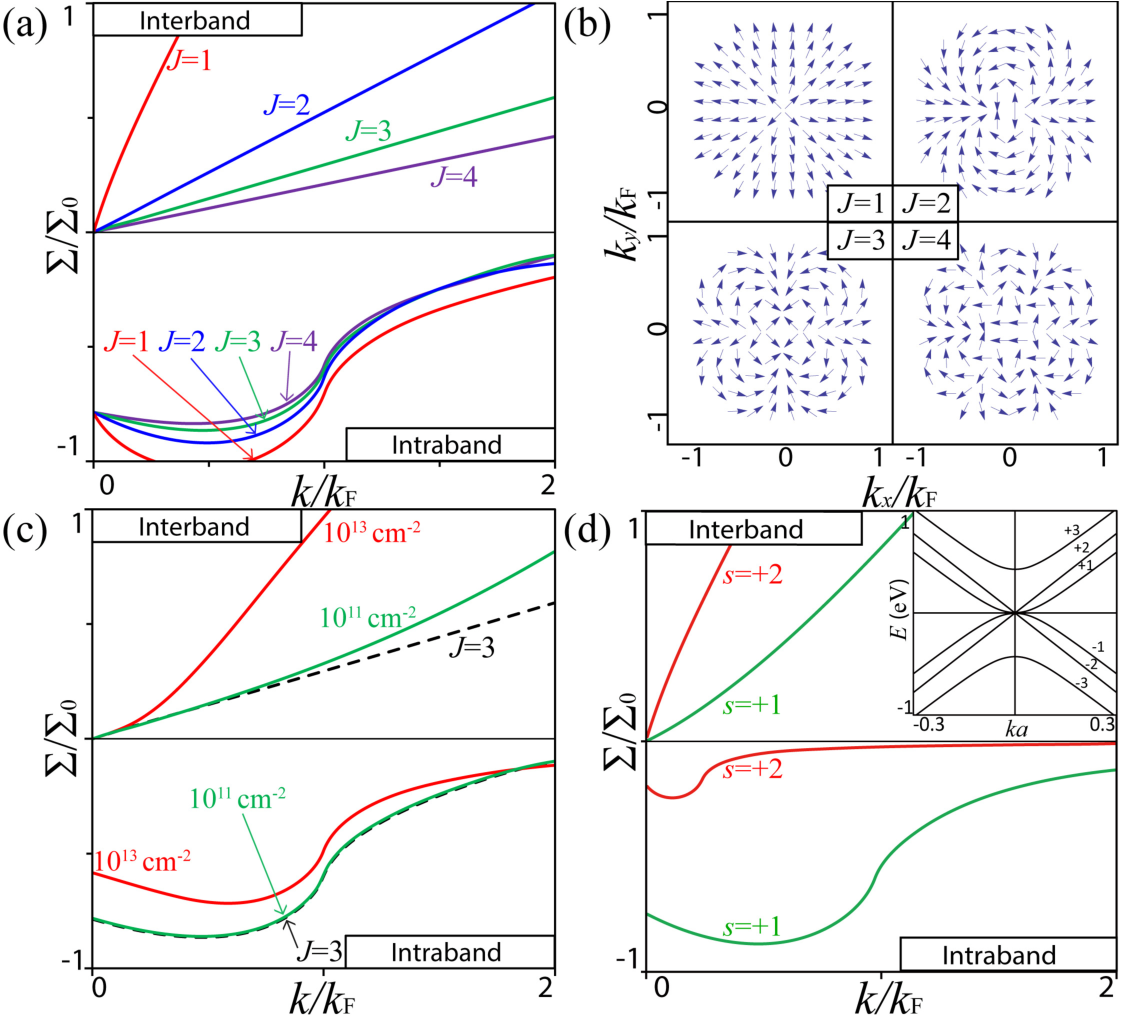}
\caption{
(a) Exchange self-energies and (b) conduction band pseudospin direction ($s=+1$) for C2DESs with $J=1,2,3,4$. 
Exchange self-energies of (c) the lowest conduction band ($s=+1$) of ABC trilayer graphene for $n=10^{11}, 10^{13}$ cm$^{-2}$, 
and (d) the lowest ($s=+1$) and second lowest ($s=+2$) conduction bands of ABA trilayer graphene for $n=10^{12}$ cm$^{-2}$.
(The bands of the ABA multilayer structures are shown in the inset.)
Here $\Sigma_0={2 e^2 k_{\rm F} \over \epsilon_0 \pi}$ and we use the effective fine structure constant $\alpha={e^2 \over \epsilon_0 \hbar v}=1$ and momentum cutoff $q_c=1/a$.
} 
\label{fig:self_energy}
\end{figure}

In Fig.~\ref{fig:self_energy}(a) intraband and interband contributions to the conduction band exchange self-energy of a 
C2DES are plotted. In Fig.~\ref{fig:self_energy}(b) pseudospin chirality is illustrated by plotting the spin-1/2 pseudospin orientation 
appropriate for two-component spinors.  
As the chirality increases, the magnitude of each contribution is suppressed because pseudospin orientation changes more rapidly with wavevector. 
Especially, the interband exchange is suppressed more strongly owing to the contribution from states occupying the infinite sea of negative energies.
In Figs.~\ref{fig:self_energy}(c) and (d) we compare 
C2DES exchange self-energies with those of ABC and ABA graphene multilayers. 
At low carrier densities in ABC graphene, the relative chiral index of the dominant wavefunction components is $3$ and 
the exchange self-energy resembles the weak form found in a C2DES with $J=3$; as the density increases, interlayer hopping becomes less important, and the exchange self-energy eventually approaches that of a C2DES with $J=1$. 
At low densities, ABA graphene is described\cite{min2008a} by a direct product of chiral gases with $J=1$ and $J=2$.

\begin{figure}[htb]
\includegraphics[width=1\linewidth]{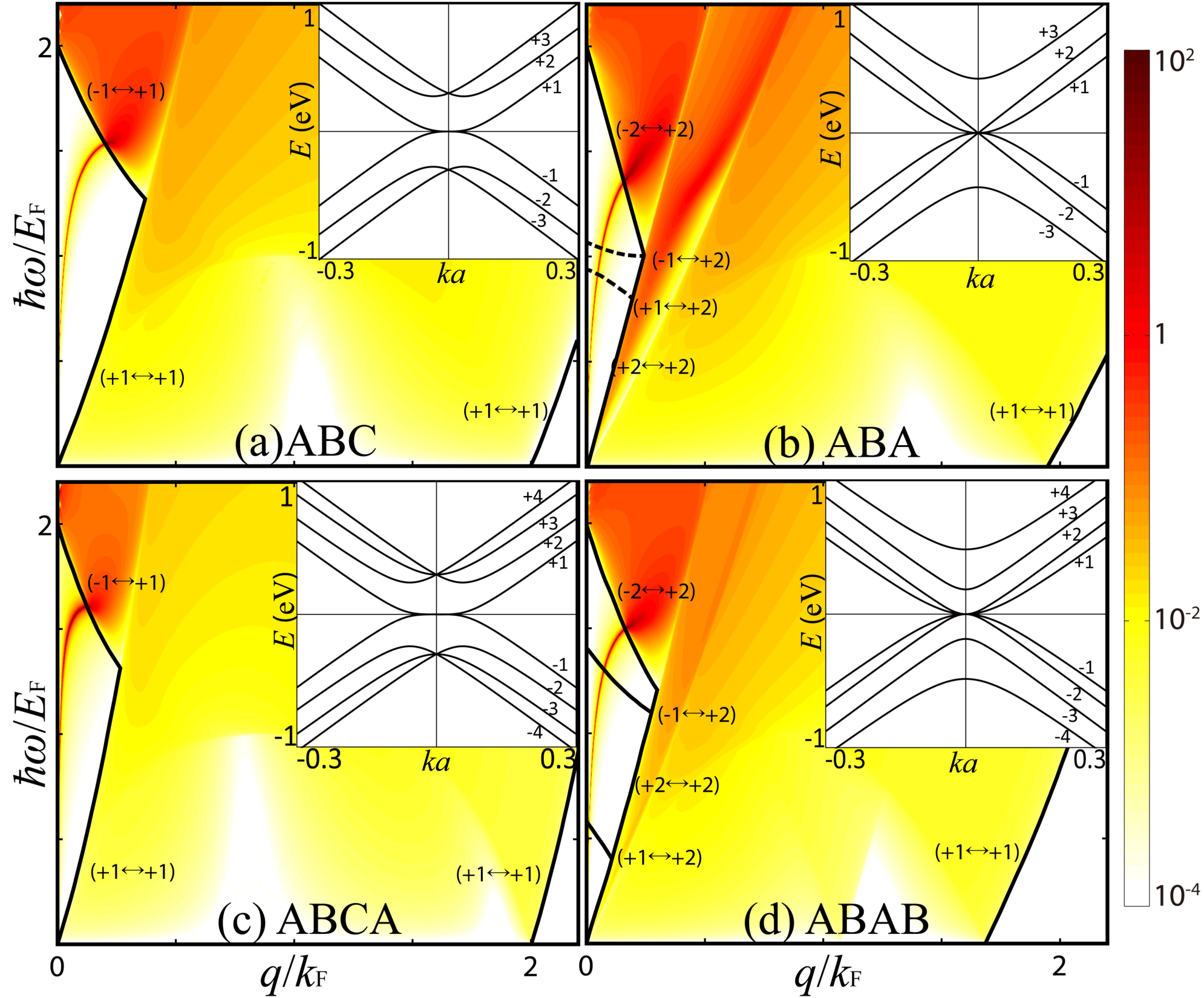}
\caption{
Loss function ${\rm Im}[-\varepsilon(q,\omega)^{-1}]$ of
(a) ABC, (b) ABA, (c) ABCA, and (d) ABAB stacked multilayer graphene for $n=10^{12}$ cm$^{-2}$ and $\alpha=1$ with $\eta=5\times 10^{-5}\varepsilon_{\rm F}$.
The thick black lines indicate boundaries of the electron--hole continua and the 
insets in each panel show the energy band structure. 
In the ABA structure and in other multilayer structures with mirror symmetry, some interband transitions do not 
contribute to plasmon Landau damping, as indicated by dotted lines in panel (b). 
} 
\label{fig:loss_function}
\end{figure}

{\em Density--density response functions and collective modes} --- 
Figure \ref{fig:loss_function} plots loss functions ${\rm Im}[-\varepsilon(q,\omega)^{-1}]$ for
several different multilayer graphene structures.  Here $\epsilon(q,\omega)$ is the dielectric function which 
we approximate using the weak coupling random phase approximation (RPA) expression
$\epsilon(q,\omega)=1-V_q \Pi_0(q,\omega)$, and 
$\Pi_0(q,\omega)$ is the non-interacting electron density--density response function: 
\begin{equation}
\Pi_0(\bm{q},\omega)=g_{\rm sv}\sum_{s,s'}\int {d^2 k \over (2\pi)^2} {f_{s,\bm{k}}-f_{s',\bm{k+q}} \over  \hbar\omega + \Delta_{\bm{k},\bm{k+q}}^{s,s'} + i \eta} 
F_{\bm{k},\bm{k+q}}^{s,s'},
\label{eq:bare_polarization}
\end{equation}
where $g_{\rm sv}=g_{\rm s} g_{\rm v} =4 $ is the spin-valley degeneracy, 
$\Delta_{\bm{k},\bm{k+q}}^{s,s'}= \varepsilon_{s,\bm{k}}-\varepsilon_{s',\bm{k+q}}$, $\varepsilon_{s,\bm{k}}$ is the eigenenergy for the band index $s$ and wavevector $\bm{k}$, and $\eta$ is a positive infinitesimal number.  
The black thick lines in Fig.~\ref{fig:loss_function} plot the boundaries of electron--hole continua within which ${\rm Im}\Pi_0({\bm q},\omega)$ is non-zero and electron--hole excitations are allowed. 
When $\epsilon(q,\omega)=0$, the loss function has a $\delta$-function peak corresponding to 
plasmon collective excitations.  When the plasmon modes enter the electron--hole continuum, 
they can decay into single electron--hole pairs through the Landau damping process. 
In multilayer graphene, plasmon modes decay through interband transitions. 
The shark-fin structures around $\omega=0$ reflects from the chiral nature of the wavefunctions 
which lead to suppressed momentum-dependent scattering\cite{borghi2009}. 


{\em Ground State Energy} --- 
The ground-state energy is the sum of the non-interacting kinetic energy and interaction (exchange-correlation) energies. 
The exchange-correlation energy can be expressed in terms of the density-density response function\cite{giuliani2005} 
by applying the integration-over-coupling-constant method and appealing to the fluctuation-dissipation theorem. 
The RPA approximation to the exchange-correlation energy is 
justified in part by the 
relatively large spin-valley flavor degeneracy $g_{\rm sv} =4$ which makes the RPA bubble-diagram 
contributions to the energy more dominant\cite{giuliani2005}.
For technical reasons it is convenient to separate the first-order exchange-correction to the interaction energy
and higher order corrections commonly referred to as the correlation energy. 
The dependence of the exchange and RPA correlation energies on carrier 
density can then be expressed\cite{barlas2007,polini2007}
as integrals along the imaginary frequency axis:
\begin{eqnarray}
\varepsilon_{\rm ex}&=&-{\hbar \over 2n} \int {d^2q \over (2\pi)^2} \int_0^{\infty} {d\nu \over \pi} V_q\delta \Pi_0(q,i\nu), \\
\varepsilon_{\rm corr}&=&{\hbar \over 2n} \int {d^2q \over (2\pi)^2} \int_0^{\infty} {d\nu \over \pi} \bigg[ V_q\delta \Pi_0(q,i\nu) \nonumber\\
&+&\ln\left| {1-V_q\Pi_0(q,i\nu) \over 1-V_q\left.\Pi_0(q,i\nu)\right|_{n=0}} \right| \bigg],\nonumber
\label{eq:exchange_correlation_energy}
\end{eqnarray}
where $\delta\Pi_0(q,i\nu)=\Pi_0(q,i\nu)-
\left.\Pi_0(q,i\nu)\right|_{n=0}$.
We use the momentum cutoff $q_{\rm c}=1/a$ to remove the ultraviolet divergences in the momentum integrals. 

Using these expressions, we find that in terms of the dimensionless coupling constant $\alpha_{\rm F}= e^2 / \epsilon_0 \hbar v_{\rm F} = (v/v_{\rm F}) \alpha$ where $v_{\rm F}$ is the Fermi velocity,   
the exchange energy is given by
\begin{equation}
\varepsilon_{\rm ex} = \hbar v_{\rm F} k_{\rm F} C_1 \alpha_{\rm F}={e^2 \over \epsilon_0} k_{\rm F} C_1,
\label{eq:exchange_coupling}
\end{equation}
and the correlation energy for small $\alpha_{\rm F}$ has the form of
\begin{equation}
\varepsilon_{\rm corr}=\hbar v_{\rm F} k_{\rm F}\left(C_2 \alpha_{\rm F}^2 + \cdots\right),
\label{eq:correlation_weak_coupling}
\end{equation}
whereas in the strong coupling limit ($\alpha_{\rm F}\gg 1$),
\begin{equation}
\varepsilon_{\rm corr}=\hbar v_{\rm F} k_{\rm F} \left(D_1 \alpha_{\rm F}+D_0 +\cdots\right).
\label{eq:correlation_strong_coupling}
\end{equation}
The coefficients $\{C_i\}$ and $\{D_i\}$ in these expressions have weak density dependence through the response function and the momentum cutoff.

\begin{figure}[htb]
\includegraphics[width=1\linewidth,height=2.65in]{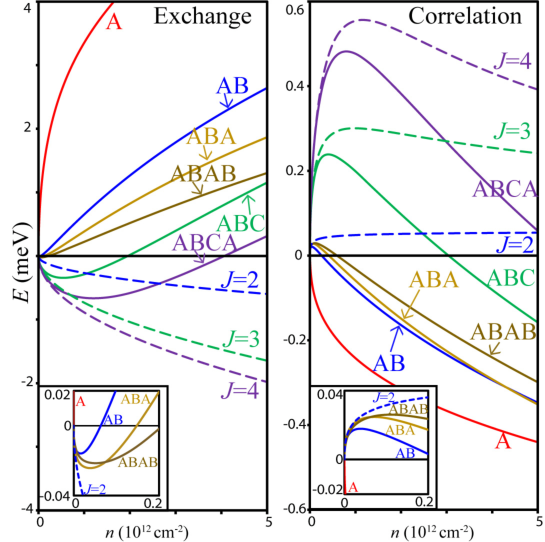}
\caption{
Comparison between the ground-state exchange energies (left panel) and correlation energies (right panel) of 
several different multilayer graphene (thick lines) structures and C2DES systems (dotted lines) as a function of carrier density for $\alpha=0.05$.
} 
\label{fig:exchange_correlation_energy}
\end{figure}
To address multilayers, we first discuss the case of C2DES models whose exchange and correlation energies exhibit a systematic dependence on their chirality index $J$. As we see from Eq.~(\ref{eq:exchange_coupling}) that the exchange energy is approximately proportional to $k_{\rm F}$ irrespective of the interaction strength.  The positive value of  $C_1$ for $J=1$ reflects the dominance in this case of interband exchange. For $J>1$, the interband contribution is suppressed because of the larger chirality indices and the exchange energy turns negative ($C_1<0$). For the correlation energy in the strong coupling limit, we find from Eq.~(\ref{eq:correlation_strong_coupling}) that $\varepsilon_{\rm corr}=(e^2 / \epsilon_0) k_{\rm F} D_1+J\varepsilon_{\rm F} D_0+\cdots$, where $\varepsilon_{\rm F}$ is the Fermi energy. Note that $C_1=-D_1$; thus for $J>1$ $\varepsilon_{\rm corr}$ is positive ($D_1>0$), whereas for $J=1$ $\varepsilon_{\rm corr}$ is negative ($D_1<0$). In the weak coupling limit, we see from Eq.~(\ref{eq:correlation_weak_coupling}) that $\varepsilon_{\rm corr}\propto v_{\rm F}k_{\rm F} a_{\rm F}^2\propto k_{\rm F}^{2-J}$.

Figure \ref{fig:exchange_correlation_energy} illustrates these properties of C2DES models and 
compares these exchange-correlation energy properties with those of multilayer graphene systems.  
For multilayer graphene, at low densities, the exchange and correlation energies follow those of the largest 
$J$ C2DES model contained within its low-energy bands because they are responsible for the largest density of states (DOS).
For example, for ABA stacking the exchange and correlation energies at low densities 
follow those of a $J=2$ C2DES because a $J=2$ C2DES has a larger DOS than a $J=1$ C2DES.
As the carrier density increases, interlayer hopping becomes less important and the exchange and correlation energies 
begin to approach those of monolayer graphene. 


{\em Summary and Discussion} --- 
In this paper we have exploited the simple dependence of 
band wavefunctions on momentum orientation to simplify many-electron
perturbation theory calculations for multilayer graphene, and to bring 
out the relationship between stacking sequence and carrier-carrier 
interaction phenomena in this interesting class of materials.
By explicit calculations for a variety of different structures we have shown that the exchange self-energies and related exchange-correlation energy features in multilayer graphene systems follow those of C2DES models at low carrier densities, 
but cross over to be more similar to those of monolayer graphene as carrier densities increase.  
The rotational transformation of the chiral wavefunction is very general and can be applied even in the presence of 
site energy variations
or remote hopping terms, unless momentum-dependent hopping terms do not appear in the Hamiltonian. For example, if the remote interlayer hopping term $\gamma_2$ is included in the Hamiltonian, it modifies $\{c_i\}$, but not the angular part of the wavefunction in Eq.~(\ref{eq:wavefunction_rotated}). 

The model we employ, however, does not include momentum-dependent remote interlayer hopping terms. 
We also use the weak-coupling RPA in the calculation.
Both limit the applicability of our calculations to moderate to high carrier densities,
and at very low carrier densities correlations frequently become strong and lead to broken symmetry 
ground states not captured by the RPA\cite{kotov2012}.

Our theory, however, captures important observable effects produced by interactions which strongly depend on the stacking sequences such as plasmon collective excitations and self-energies.
The dependence of ground state energies on carrier and spin densities 
are responsible for renormalized electronic compressibility and spin susceptibility, respectively. 
For example, the electronic compressibility $\kappa$ is given by $\kappa^{-1}=n^2 d\mu/dn$, 
where $\mu=\partial (n\varepsilon_{\rm tot})/ \partial n$ is the chemical potential of the interacting system, and $\varepsilon_{\rm tot}$ is the total  ground-state energy per particle.   
In an ordinary parabolic band two-dimensional electron system the compressibility 
famously becomes negative at low carrier densities\cite{eisenstein1992}.
In contrast, for a C2DES we find that in the low-density strong-coupling limit, 
$\kappa_0/\kappa\approx 1+J(J+2)D_0/2$ with $D_0>0$ for $J=1$ and $D_0<0$ for $J>1$.  (Here $\kappa_0$ is the non-interacting compressibility.)
It follows that for a $J=1$ C2DES, the electronic compressibility is strongly suppressed by interactions due to the interband exchange contribution and remains positive\cite{barlas2007}. 
For a $J>1$ C2DES, the interband exchange contribution is suppressed with the chirality; thus, the electronic compressibility is enhanced by the interactions. 
Interestingly, we find that for $J\geq 5$, the compressibility can be negative in the low-density limit, suggesting an instability toward other ground states, whereas in the absence of correlation, this occurs for $J\geq 2$.  
For multilayer graphene, at low carrier densities, the compressibility follows the trend of the corresponding C2DES, but as the carrier density increases, the compressibility follows that of monolayer graphene with suppressed compressibility\cite{kusminskiy2008,borghi2010}, showing non-monotonic behavior arising from competition between the intraband exchange, interband exchange, and correlation.

In conclusion, our new approach allows us to effectively calculate the quasiparticle and thermodynamic properties of interacting many-body chiral systems. 
We show that as the chirality increases, the exchange contribution to the single particle energy is suppressed and the correlation contribution increases, indicating that the exchange-correlation is controlled by the stacking arrangement. 
Our results suggest that correlation effects play a more important role in a system with a large chirality; thus, we expect that rhombohedral graphene with periodic ABC stacking could show exotic interaction-induced phenomena such as ordered states and non-Fermi liquid behavior\cite{kotov2012}. 
%


\acknowledgments 
This research was supported by the Basic Science Research Program through the National Research Foundation of Korea (NRF) funded by the Ministry of Science, ICT and Future Planning under Grant No. 2012R1A1A1013963 (YJ and HM), Basic Science Research Program 2009-0083540 (EHH) and DOE Materials Science and Engineering grant DE-FG03-02ER45958 and by Welch Foundation grant TBF1473 (AHM). HM thanks Eun-Gook Moon for helpful discussions.


\end{document}